\documentclass[twocolumn,samsmath,amssymb,floatfix,superscriptaddress,nofootinbib]{revtex4}
\usepackage{graphicx}
\begin{document}
\title{
Equilibrium cluster phases and low-density arrested disordered states:  
The role of short-range attraction and long-range repulsion} 
\def\roma{\affiliation{
Dipartimento di Fisica and INFM Udr and Center for
Statistical Mechanics and Complexity, Universit\`a 
di Roma ``La Sapienza'', P.le A. Moro 2, I-00185, Roma, 
Italy}}
\def\esrf{\affiliation{
European Synchrotron Radiation Facility. B.P.{\em 220} F-{\em 38043} Grenoble,
Cedex France.
}}
\author{Francesco Sciortino}\roma
\author{Stefano Mossa}\roma,\esrf
\author{Emanuela Zaccarelli}\roma
\author{Piero Tartaglia}\roma
\date{\today}
\begin{abstract}
We study a model in which particles interact with short-ranged
attractive and long-ranged repulsive interactions, in an attempt to
model the equilibrium cluster phase recently discovered in sterically
stabilized colloidal systems in the presence of depletion
interactions.  At low packing fraction particles form stable
equilibrium clusters which act as building blocks of a cluster fluid.
We study the possibility that cluster fluids generate a low-density
disordered arrested phase, a gel, via a glass transition driven by the
repulsive interaction.  In this model the gel formation is formally
described with the same physics of the glass formation.
\end{abstract}
\maketitle
Gels  are disordered arrested systems capable of sustaining 
weak shear stress. Although  commonly used in food, cosmetics and 
detergents, the physical processes involved in the transformation 
of a solution capable of flowing into a highly viscous substance, 
and eventually into a gel state, are far from being understood.  
The gel state can be generated via several routes. 
Interrupted phase separation, bond percolation, irreversible diffusion 
limited aggregation, and attractive glass transition have been suggested 
as possible mechanisms for gel formation in polymeric and colloidal systems.  

Sterically stabilized colloidal systems have often been used as an
experimental realization of the hard-sphere model, to study both
fluid-crystal transition and glass formation. The addition of small
polymers in solution modifies the effective colloid-colloid
interaction by depletion mechanisms, complementing the hard-sphere
repulsion with a short-range attraction, on a length scale of the
order of a few percent of the colloid size.  

At small volume fractions $\phi$ and low temperature $T$, the systems
experience dynamic arrest in a transient gel structure~\cite{poon97,lekker-exp-gel}.  
Recent experiments suggest that the gel region in the ($\phi-T$) plane 
is contiguous to a cluster phase, in which 
particles~\cite{prlweitz,daveweitz,poon-exp-gel} form
freely diffusing clusters of finite size.  Studies of the phase
diagram of short-ranged (uncharged) attractive colloidal systems show
that these systems are characterized by fluid and crystal phases and
by a very flat metastable liquid-gas spinodal, which extends to
considerably low values of
$\phi$~\cite{lekker,lekkerkerker02,frenkeltermo}. In simulations of
short-ranged attractive potentials, gel phases have only been observed
as a result of an arrested phase separation, but with no contiguity
with an equilibrium cluster phase~\cite{capri}.

In this Letter we provide evidence, via molecular dynamics (MD)
simulations, that when short-range attractive interactions are
complemented by long-range repulsive interactions, equilibrium cluster
phases, as well as disordered arrested states contiguous to cluster
phases, can be generated.  The long-range repulsive interactions may
arise from screened electrostatic interactions or from effective
interactions accounting for the presence of co-solutes in
solution~\cite{likos,loewen}.

\begin{figure}[t]
\centering
\includegraphics[width=0.42\textwidth]{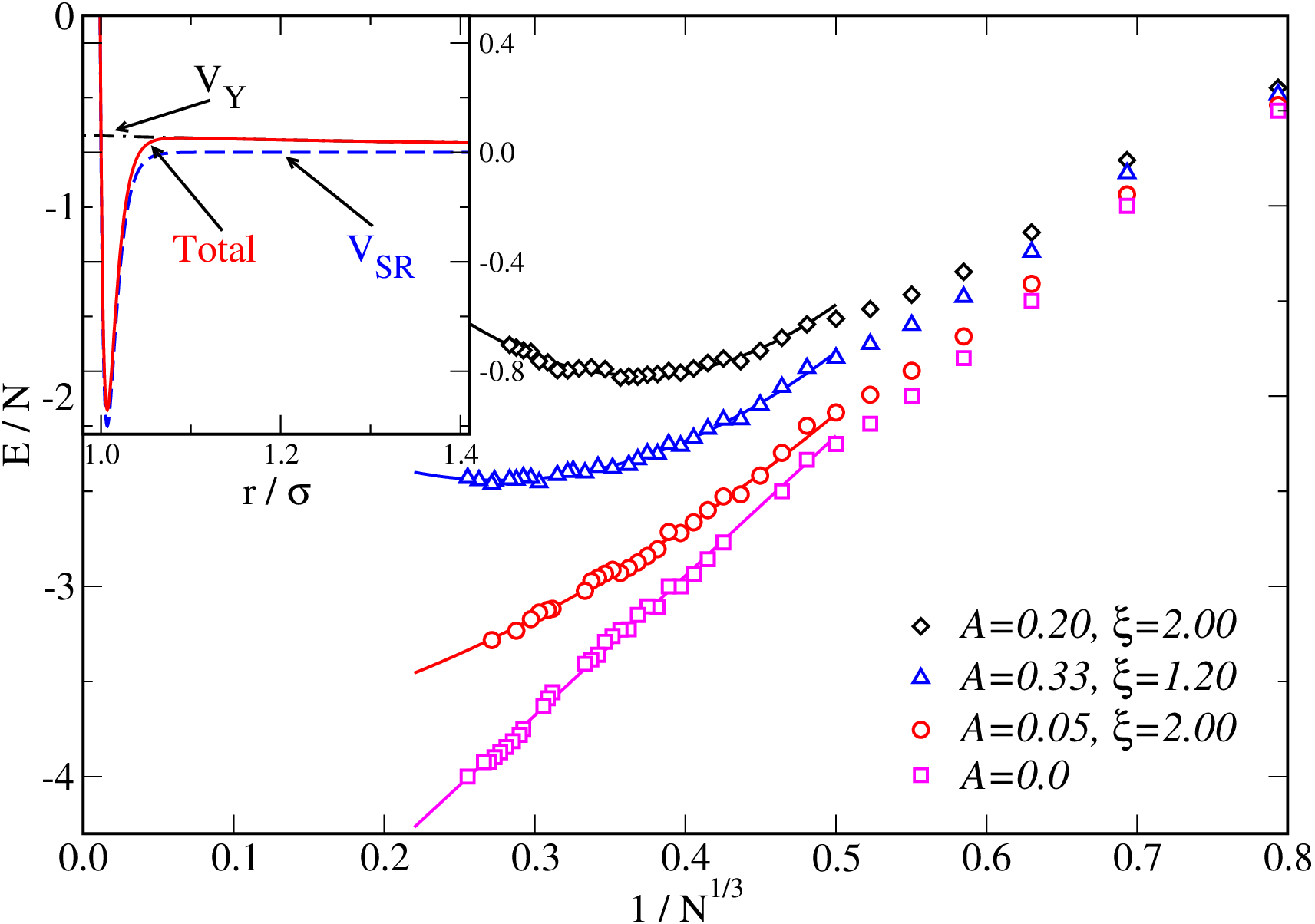}
\caption{Ground state energy for connected clusters composed of $N$
particles for different choices of $A$ and $\xi$, with $\alpha=100$.
While in the $A=0$ case, a monotonous decrease of the energy ---
linear in $N^{-1/3}$ as expected from the balance between bulk and
surface effects --- is found, for large $A$ the lowest energy state is
reached for a finite $N^*$ value, which provides an upper limit to the
size of thermodynamically stable clusters. The inset shows $V_{SR}$,
$V_{Y}$, and their sum for the case $A=0.05$ and $\xi=2$.}
\label{fig:energy}
\end{figure}

The possibility of generating a cluster phase, i.e., an equilibrium
state in which colloidal particles partition into stable clusters via
the competition between attraction (which favors cluster growth at low
$T$) and long-range repulsion (which favors low local particle
densities and, therefore, small aggregates) has been discussed in
previous theoretical works~\cite{chandler,xxxx,kegel,muratov}.  
Here we show that in the presence of an appropriate balance between
attraction and repulsion the system forms stable clusters, around an
optimal cluster size. To quantify this proposition we calculate the
ground state energy of clusters of different size\cite{noi,wales97}
for the case of particles interacting with a potential $V(r)$ which is
the sum of a short-range attractive potential $V_{SR}$ (modeled for
simplicity with the generalization to very large $\alpha$ of the
Lennard-Jones $2\alpha-\alpha$ potential, as proposed by Vliegenthart {\em et
al.}~\cite{lekker})
\begin{equation} 
V_{SR}(r)=4 \epsilon  \bigg[ \bigg(\frac{\sigma}{r}\bigg)^{2\alpha}- 
\bigg(\frac {\sigma}{r}\bigg)^{\alpha}\bigg], 
\label{eq:potsr}
\end{equation}  
and of a screened electrostatic repulsive interaction $V_Y$ 
(modeled by a Yukawa potential) 
\begin{equation}
V_{Y}(r)=A\; \frac{e^{-r/\xi}}{r/\xi}.
\label{eq:potyuk}
\end{equation}
In the following, the parameters $\epsilon$ and $\sigma$ entering in
$V_{SR}$ are chosen as units of energy and distance. The parameter $\alpha$
has been chosen equal to $100$, corresponding to an attractive range
of about three percent of $\sigma$ (see inset in Fig.\ref{fig:energy}). 
For this value of $\alpha$, the liquid-gas critical
temperature for $V_{SR}$ is found to be $0.235 \pm0.005$. 
Fig.~\ref{fig:energy} shows that the cluster ground state
energy, for particles interacting with the potential $V_{SR}+V_{Y}$
for appropriate values of $A$ and $\xi$ has a minimum at a finite size
$N^*$, indicating that clusters of size larger than $N^*$ are
energetically disfavored (and, hence, that liquid condensation is
inhibited). Since entropic contributions to the free energy will
always favor small clusters, aggregation will never proceed beyond
$N^*$.  The existence of an optimal cluster size (which will of course
depend on $T$ and $\phi$) suggests that, at low $T$, clusters act as
building blocks of the supra-molecular cluster fluid.

The cluster-cluster interaction is essentially controlled by the
Yukawa repulsion, due to the short-range nature of the attractive part
of the potential. It can be calculated analytically, if one
approximates the cluster to a sphere of radius $R$ composed of a
homogeneous density $\rho$ of points interacting via the potential
$V_Y$ (Eq.~(\ref{eq:potyuk})). The resulting cluster-cluster
interaction is, again, of the Yukawa type with the same $\xi$ but a
renormalized amplitude $A(R)$
\begin{equation}
{{A(R)}\over{A}}=
\left\{  2\pi \xi^3 \rho \, e^{-R/\xi} \left[ 1+{{R}\over{\xi}}
+  \left({{R}\over{\xi}}-1\right) 
e^{{2R}/{\xi}} 
\right]\right\}^2.
\label{eq:flowA}
\end{equation}
Therefore, aggregation of particles into clusters of
size $N$ (and radius $R \sim N^{1/3}$) has two effects: 
{\em i)} renormalization of the amplitude of the interaction, 
and {\em ii)} decrease of the cluster number density $n$ according to
\begin{equation}
{{n(R)}\over{n}}={{1}\over{N}}.
\label{eq:flowphi}
\end{equation}
Since $\xi$ is not modified, the aggregation process can be seen as
the generator of a flow, defined via Eqs.~(\ref{eq:flowA}) and
(\ref{eq:flowphi}), in the phase diagram of the pure repulsive Yukawa
potential.

\begin{figure}[h]
\centering
\vspace{0.2cm}
 \includegraphics[width=0.40\textwidth]{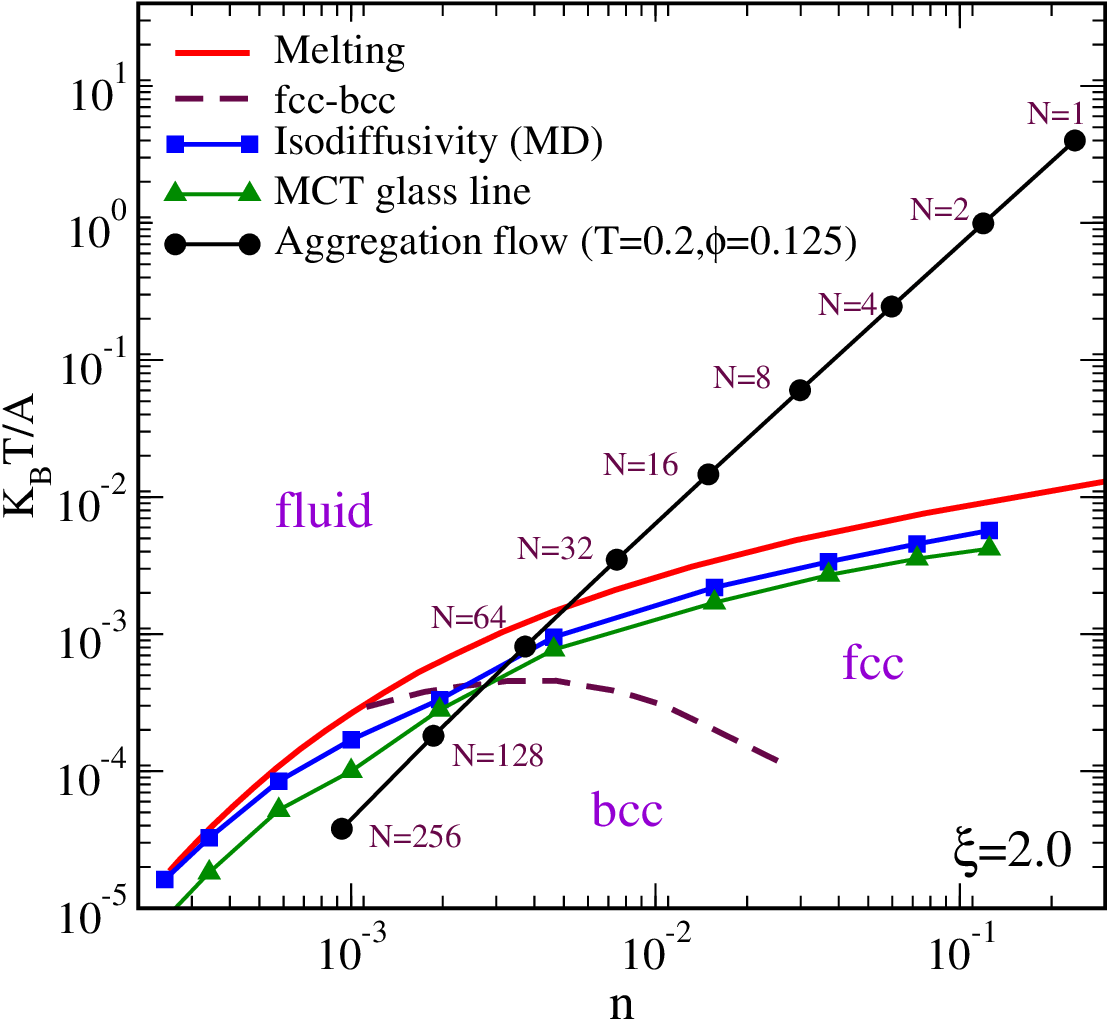}
 \vspace{0.2cm}
 \caption{Extended phase diagram for the Yukawa potential, for
$\xi=2$, reporting the fluid-crystal phase boundary
lines~\protect\cite{grest}, the iso-diffusivity locus ($D =10^{-4} $
in unit of $\sqrt{A \xi^2/m}$) and the ideal MCT glass line.  The flow
line, calculated according to Eqs.~(\ref{eq:flowA}) and
(\ref{eq:flowphi}), describes a system of Yukawa particles ($A=0.05,
\xi=2$), with packing fraction $\phi=\pi/6 \sigma^3 n=0.125$ and
$T=0.20$, assumed to progressively form spherical homogeneous clusters
of size $1$, $2$, $4$, $8$,$\ldots$, $128$. When the cluster size
reaches $N \approx 47$, the melting line is crossed and the fluid
transforms into a fcc-crystal whose elementary units are clusters of
that size.  If crystallization is avoided, the ideal MCT glass line is
crossed when the average cluster size reaches $N \approx 86$. These
figures are in very good agreement with the simulation data shown in
Fig.~\ref{fig:cluster}.  }
\label{fig:flow}
\end{figure}

Fig.~\ref{fig:flow} shows an extended phase diagram for the pure
Yukawa potential. Together with the fluid-crystal coexistence lines,
previously calculated by Kremer {\em et al.}~\cite{grest}, and Meijer
and Frenkel~\cite{frenkel}, we show an iso-diffusivity
curve\cite{simyuk} (i.e., the locus of point with the same diffusion
constant) and the ideal glass line, evaluated within the Mode Coupling
Theory~\cite{mct} (MCT) using {\em exact} numerical structure factors
as input. The latter curve provides an upper limit to the glass
transition line.  It is important to note that the long-range nature
of the repulsion enables the formation --at low $k_B T/A$ values-- of
solid phases with extremely low density (named Wigner crystal or
Wigner glass~\cite{bosse,lai97,laponite}).

An example of the flow in the Yukawa phase diagram induced by
clustering is shown in Fig.~\ref{fig:flow}. The starting point is
assumed to be a system composed of monomers ($N=1$) interacting via
the potential $V_{SR}+V_{Y}$.  If $T$ is sufficiently low, a
clustering phenomenon will take place, driven by the attractive part
of the potential. Such aggregation process acts as a generator of a
flow in the phase diagram, moving the system, due to the effective
increase of the repulsive amplitude (Eq.~\ref{eq:flowA}) and to the
associated decrease of the cluster number density
(Eq.~\ref{eq:flowphi}), closer and closer to the melting and glass
lines.  If the optimal size is reached before crossing the melting
line, the equilibrium state for the system will correspond to a fluid
phase of clusters. In the other case, an arrest phenomenon will take
place, driven by the crystallization or --more likely because of the
inherent self-generated cluster polydispersity-- by a glass
transition. With this mechanism arrested disordered states with small
$\phi$ can be generated. This new route to the gel state demonstrates,
in the case of weakly charged short-ranged attractive colloids, the
identity of the gel and the glass states of matter.

\begin{figure}[t]
\centering
\includegraphics[width=0.45\textwidth]{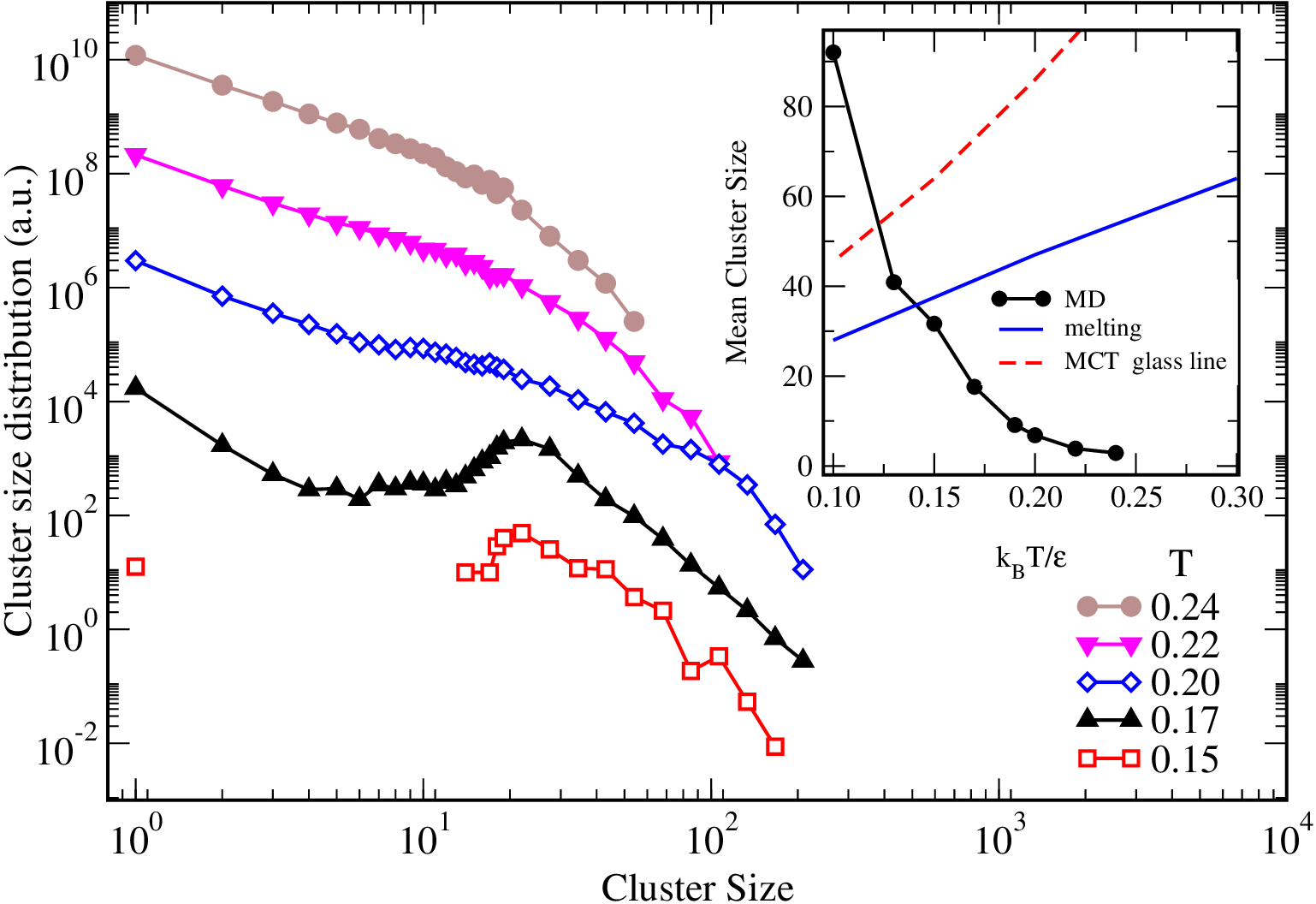}
\caption{Cluster size distribution, on a log-log scale, for
$\phi=0.125$. For clarity, curves for different $T$ have been
shifted along the vertical axis progressively by one decade.
The inset shows the $T$-dependence of the average cluster size. It also shows, for each
$T$, the size at which the crystallization and the ideal glass
transition lines are crossed, according to the flow described in
Fig.~\protect\ref{fig:flow}.  }
\label{fig:cluster}
\end{figure}

We next provide numerical evidence that the above proposed mechanism
does lead to disordered arrested states contiguous to a cluster phase.
We perform MD simulations of particles interacting with the potential
$V_{SR}+V_{Y}$\cite{simudetails}, both along a constant $T$ and along
a constant $\phi$ path. Here, we focus on the specific choice of
$A=0.05$ and $\xi=2$, referring to future publications
for a more extensive analysis. This choice of parameters is
characteristic of colloidal particles in weakly polar
solvents~\cite{daveweitz,bladereen,grier}.

The cluster size distribution, shown in Fig.~\ref{fig:cluster},
provides a characterization of the clustering process. While at high
$T$ the cluster size is monotonically decreasing, around $k_B T/
\epsilon=0.2$ the distributions develop a peak at a finite value,
signaling the presence of stable equilibrium aggregates composed of
$20-30$ particles.  We note that such resulting cluster phase can be
considered as an echo of the interrupted spinodal decomposition, since
clustering takes places just below the liquid-gas critical temperature
of the bare $V_{SR}$ potential.

For each of the studied $T$, the unclustered system ($N=1$) can be
represented as a point in Fig.~\ref{fig:flow}.  A flow curve, starting
from each of these points, can be calculated according to
Eqs.~\ref{eq:flowA} and \ref{eq:flowphi}. In this way, a theoretical
prediction for the average cluster size at which crystallization and
(MCT) vitrification are expected can be derived. Such curves are shown
in the inset together with the $T$-dependence of the average cluster
size calculated from the MD data.  For the present choice of $A$,
$\xi$ and $\phi$, structural arrest (either in the form of
crystallization or in the form of vitrification) is predicted when
the average cluster size is  $\gtrsim35$ particles ($k_BT/\epsilon \leq 0.15$).

\begin{figure}[t]
\centering
\includegraphics[width=0.35\textwidth]{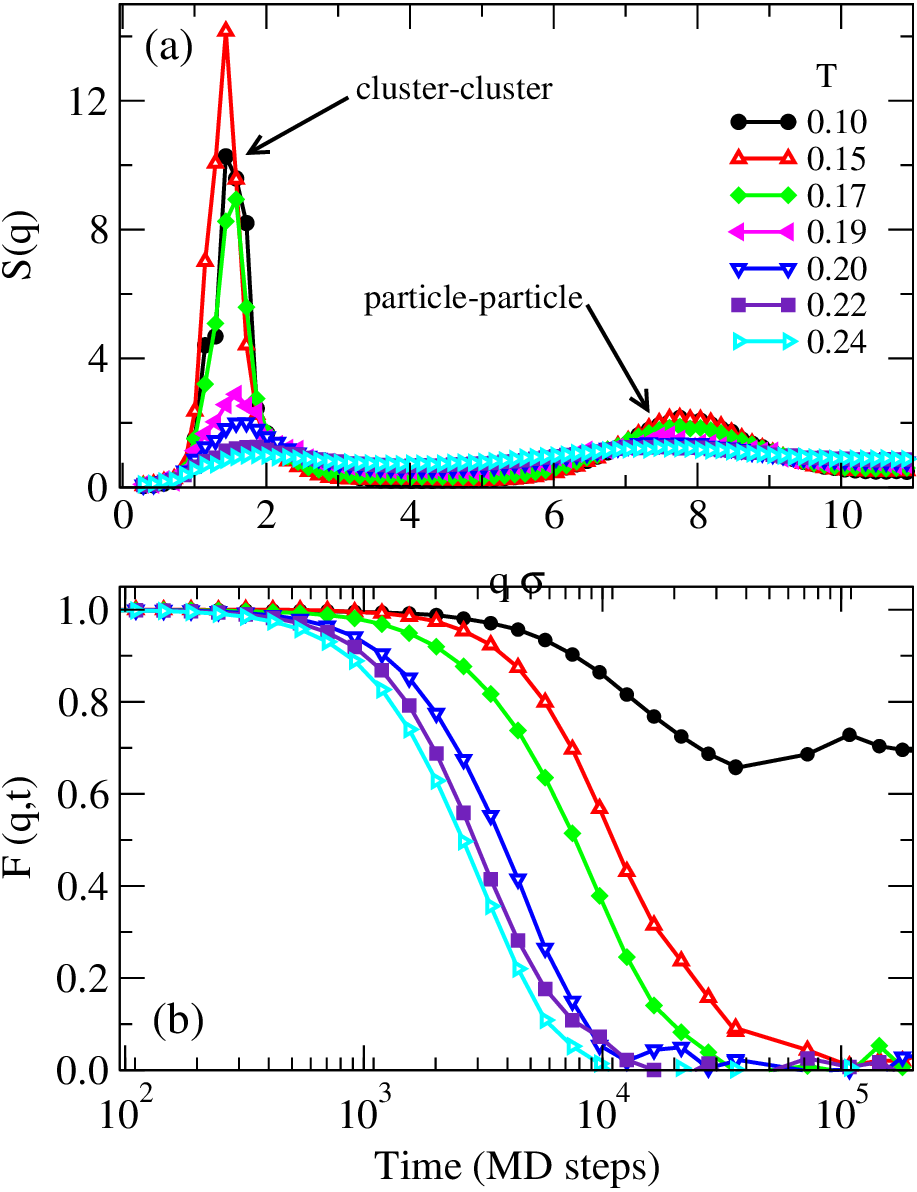}
\caption{ Evolution of static and dynamic quantities on cooling at
$\phi=0.125$. (Top) $S(q)$. Note the development of
a peak at small $q$, reflecting the formation of clusters in the
system. (Bottom) $F(q,t)$ at $q\sigma=2.7$.  At $T = 0.1$ the 
autocorrelation function does not decay to zero any longer, 
highlighting the formation of a structurally arrested phase. 
}
\label{fig:sq}
\end{figure}

Before turning to the evaluation of dynamical properties, we discuss
the evolution of the structure with $T$, analyzing the static
structure factor $S(q)$. As shown in Figure~\ref{fig:sq}-a, $S(q)$
progressively develops a peak at wave-vectors $q$ associated with
distances in real space of the order of several particle diameters,
reflecting the nearest neighbor cluster-cluster distance. The
amplitude of this novel peak changes significantly in a small
temperature range around the $T$-region where the flow diagram
predicts the crossing of the crystallization line.  Therefore, the
system changes from a solution of interacting particles toward a
system of larger and larger interacting clusters.  In the cluster
phase, the location of the peak changes very weakly both with $T$ and
with $\phi$. It is interesting to note that both the location and the
amplitude of the cluster-cluster peak are in agreement with the
experimental data for colloidal gels reported in Ref.~\cite{prlweitz}.
To confirm that the aggregation process indeed generates a disordered
arrested state, we show in Fig.~\ref{fig:sq}-(Bottom) the coherent
scattering function $F(q,t)$. It displays a (one-step) decay, with a
characteristic time which increases on cooling. At the lowest studied
$T$, the correlation function does not decay to zero, signaling that
the system is trapped in a non-ergodic state. The transition from
fluid to glass is very sharp, again in full agreement with results for
colloidal gels~\cite{prlweitz}.

In summary, we have proposed and studied a model which shows
equilibrium cluster phases and arrested states at low densities,
opening the way for accurate comparisons between experiments and
theoretical predictions.  Contrary to what has been proposed before
for attractive colloids, we suggest that, in these systems, the gel
phase is not stabilized by the short-range attraction (and hence its
formation is not connected to the attractive glass transition observed
at high packing fractions~\cite{fuchs}).  The simulation data
(Figs.~\ref{fig:cluster} and \ref{fig:sq}) show that the competition
between isotropic short-range attraction and long-range repulsion ---
of the type found in colloidal systems and in protein
solutions~\cite{piazza00,pellicane03} --- is capable of generating a
very rich range of phenomena~\cite{douglas}.  Within the present model,
the formation of low density arrested states can be modeled as a glass
transition, where clusters (as opposed to particles) are trapped in
cages generated by the long-range repulsions.  It will be interesting
to study if the arrested disordered state is metastable with respect to a a
fcc or bcc crystal of "spherical" clusters (a Wigner crystal) or with
respect to lamellar phases, which has been observed in related
models~\cite{chandler,xxxx,muratov}.

\noindent We thank D. Chandler, D. Frenkel, W. Kegel, D.R. Reichman,
G. Ruocco, and I. Saika-Voivod for comments. F. S. thanks W. Poon for 
suggesting to investigate the role of long-range repulsive interactions 
in short-range attractive colloidal systems. Authors acknowledge support
from MIUR Cofin 2002 and Firb.

\begin{thebibliography}{99}
%
\bibitem{poon97} 
W.~C.~Poon {\it et al.}, Physica A {\bf 235}, 110 (1997).
%
\bibitem{lekker-exp-gel}
N.~A.~M. Verhaegh  {\it et al.}, Physica A {\bf 264}, 64 (1999).
%
\bibitem{prlweitz}
P.~N Segr\`e {\it et al.}, Phys. Rev. Lett. {\bf 86}, 6042 (2001).
%
\bibitem{daveweitz} 
A.~D. Dinsmore  and D.~A. Weitz, J. Phys.: Condens. Matter {\bf 14}, 7581 (2002).
%
\bibitem{poon-exp-gel} 
H. Segdwick  {\it et al.}, preprint cond-mat/0309616 (2003).
%
\bibitem{lekker} 
G.~A. Vliegenthart {\it et al.}, Physica A {\bf 263}, 378 (1999).
%
\bibitem{lekkerkerker02}
V.~J. Anderson  and H.~N.~W. Lekkerkerker, Nature {\bf 416}, 811 (2002). 
%
\bibitem{frenkeltermo} 
M.~A. Miller and D. Frenkel, Phys. Rev. Lett. {\bf 90}, 135702 (2003).
%
\bibitem{capri} 
E. Zaccarelli {\it et al.}, preprint cond-mat/0310765 (2003).
%
\bibitem{loewen} 
A.~A. Louis {\it et al.}, Phys. Rev. E  {\bf 65}, 061407 (2002).
%
\bibitem{likos} C. N.  Likos, Phys. Rep.  {\bf 348}, 267 (2002).
%
\bibitem{chandler} 
D. Wu {\it et al.}, J. Phys. Chem. {\bf 96}, 4077 (1992).
%
\bibitem{xxxx} 
R.~P. Sear {\it et al.}, Phys. Rev. E {\bf 59}, R6255 (1999).
%
\bibitem{kegel} 
J. Groenewold and W.~K. Kegel, J. Phys. Chem. B {\bf 105}, 11702 (2001).
%
\bibitem{muratov} C.~B. Muratov, Phys. Rev. E {\bf 66}, 066108 (2002).
%
\bibitem{wales97}
D.~J. Wales  and  J.~P.~K. Doye, J. Phys. Chem. A {\bf 101}, 5111 (1997).
%
\bibitem{noi} 
To evaluate the lowest energy minima, we have
implemented a modified version of the basin-hopping algorithm
introduced by Wales and Doye~\protect\cite{wales97}.  The algorithm
consists of a constant-$T$ Monte-Carlo (MC) simulation where the
acceptance criterion is based upon the energy of the closest minimum
of the potential energy surface. To favour barrier crossing, every
$100$ MC steps the least bounded atom is removed and re-inserted in
the position with lowest insertion energy.  In the search procedure
moves creating disconnected clusters are rejected.  Indeed, for $N
\gtrsim N^*$, the ground state consists of separated non-interacting
clusters.
%
\bibitem{grest} 
M.~O. Robbins   {\it et al.}, J. Chem. Phys. {\bf 88}, 3286 (1988).
%
\bibitem{frenkel}
E.~J. Meijer and D. Frenkel, J. Chem. Phys. {\bf 94}, 2269 (1991).
%
\bibitem{simyuk} 
MD simulation of 1000 particles of mass $m$
interacting with the Yukawa potential (Eq.~(\ref{eq:potyuk})) are
performed for $10$ different number densities, to evaluate the
$T$-dependence of $D$ and $S(q)$.  Iso-$D$ curves for $D<10^{-4}
\sqrt{A\xi^2/m}$ cannot be calculated, since crystallization takes
place.  The MCT equations~\cite{mct} are solved with standard
techniques on a grid of $250$ wave-vectors, to evaluate the ideal
glass transition line.
%
\bibitem{mct}  
W. G\"otze, in {\it Liquids, Freezing and Glass Transition}, p.287,  
Hansen, J.~P., Levesque, D. and  Zinn-Justin J.(eds.),
North Holland, Amsterdam (1991).
%
\bibitem{lai97}
S.~K. Lai  {\it et al.}, Phys. Rev. E {\bf 56}, 766 (1997).
%
\bibitem{bosse} 
J. Bosse and S.~D. Wilke, Phys. Rev. Lett. {\bf 80}, 1260 (1998).
%
\bibitem{laponite}
D. Bonn {\it et al.}, Europhys. Letters {\bf 45}, 52 (1999).
%
\bibitem{simudetails} 
Data reported in Figs.~\ref{fig:cluster}-\ref{fig:sq} refer to a system 
of $N_o = 2500$ particles of unit mass $m$, interacting via the potential
$V=V_Y+V_{SR}$ defined in Eqs.~(\ref{eq:potsr}) and (\ref{eq:potyuk}).
as units of energy and distance. 
The parameters of $V_{Y}$ are $A=0.05$ and $\xi=2$. The potential
has been cut off at $8\; \sigma$; such a large cut-off is necessary to
properly account for the long-range nature of $V_{Y}$. Time is
measured in units of $\sqrt{m \sigma^2 / \epsilon}$. The integration
time step is $4.55 \times 10^{-4}$ time units. The equilibration
dynamics, performed in the NVT ensemble, is extremely slow, and often
requires more than $10^8$ integration time steps. In all cases, MD
runs were performed starting from equilibrium configurations at $k_B
T/\epsilon = 1$. At very low $T$, equilibrium is not reached and
the system remains trapped in an arrested state.  After equilibration,
NVE runs are performed to evaluate $F(q,t)$, defined as $ \langle
\rho^*({\bf q},t) \rho({\bf q},0) \rangle/S(q)$, where $\rho({\bf
q},t)=\frac{1}{\sqrt{N_o}} \sum_i e^{i{\bf q} {\bf r}^i(t)}$, ${\bf
r}^i$ are the coordinates of particle $i$ at time $t$, and $S(q)
\equiv \langle|\rho({\bf q},0)|^2\rangle$. To calculate cluster size properties a bond between two
particles is defined when the relative distance is less than $1.05\sigma$ (value
at which $V_{SR}=-0.03 \epsilon$).
%
\bibitem{bladereen} 
A. Yethiraj and A. van Blaaderen, Nature {\bf 421}, 513 (2003).
%
\bibitem{grier}  
J.~C. Crocker and D.~G. Grier, Phys. Rev. Lett. {\bf 73}, 352 (1994).
%
\bibitem{fuchs} 
J. Bergenholtz {\it et al.}, Langmuir {\bf 19}, 4493 (2003);
J. Bergenholtz and M. Fuchs Phys. Rev. E {\bf 59}, 5706 (1999).
%
\bibitem{piazza00}
R. Piazza, Curr. Opin. Colloid Interface Sci. {\bf 5}, 38 (2000).
%
\bibitem{pellicane03}
Pellicane G. {\it et al.}, J. Phys.: Condens. Matter {\bf 15}, 375 (2003).
%
\bibitem{douglas}
J. Dudowicz {\it et al.}, Phys. Rev. Lett. {\bf 92}, 045502 (2004) 
and references therein.
%
\end{thebibliography}
\end{document}